\documentclass[twocolumn,showpacs,preprintnumbers,amsmath,amssymb]{revtex4}
\usepackage{graphicx}
\usepackage{dcolumn}
\usepackage{bm}

\begin{document}

\title{Parameter study of the diamagnetic relativistic
       pulse accelerator (DRPA) in slab geometry I:
       Dependence on initial frequency ratio and
       slab width}

\author{Kazumi Nishimura$^{a)}$ and Edison Liang$^{b)}$}

\address{${}^{a)}$Los Alamos National Laboratory,
Los Alamos, NM 87545 \\
${}^{b)}$Rice University,
Houston, TX 77005-1892}

\date{\today}
%
\begin{abstract}
%
Two-and-a-half-dimensional particle-in-cell plasma
simulations are used to study the particle energization
in expanding magnetized electron-positron plasmas
with slab geometry.
When the magnetized relativistic plasma with high
temperature (initial electron and positron temperature are
$k_{B}T_{e}=k_{B}T_{p}=5MeV$) is expanding into a vacuum,
the electromagnetic (EM) pulse with large amplitude
is formed and the surface plasma particles
are efficiently accelerated in the
forward
direction owing to the energy conversion from the EM
field to the plasma particles.
We find that the behavior of the DRPA (Diamagnetic
Relativistic Pulse Accelerator) depends
strongly on the ratio of
the electron plasma frequency to the cyclotron frequency
$\omega_{pe}/\Omega_{e}$ and the initial plasma
thickness.
In the high $\omega_{pe}/\Omega_{e}$ case, the EM pulse
is rapidly damped and the plasma diffuses uniformly
without forming density peaks because the initial thermal
energy of the plasma is much larger than the field
energy.
On the contrary, in the low $\omega_{pe}/\Omega_{e}$
case, the field energy becomes large enough to energize
all the
plasma particles, which are confined in the EM pulse and
efficiently
accelerated to ultrarelativistic energies. We also find
that a thicker initial plasma increases the maximum
energy of the accelerated particles.

\end{abstract}
\maketitle

\section{Introduction}
\label{sec:intr}

In astrophysical, solar and space plasmas,
the acceleration of high energy
particles is
an intriguing and unsolved problem.
Magnetic reconnection, wave turbulence and collisionless
shocks are
often invoked for the acceleration of plasma
particles, from cosmic rays, solar flares to
gamma-ray bursts\cite{pira00} and astrophysical
jets.

A new mechanism of particle energization by freely
expanding, hot, strongly
magnetized collisionless plasmas was recently
discovered by Liang {\it et al.}
using two-and-a-half dimensional
(2-1/2D) particle-in-cell (PIC) simulations
\cite{lian03}.
This mechanism, called the diamagnetic
relativistic
pulse accelerator (DRPA), is able to convert most of the
initial magnetic energy into the ultra-relativistic
directed
kinetic energy of a small fraction of the surface
particles.  Moreover,
Liang and Nishimura\cite{lian032} have
found 
that the late-time plasma pulse of the DRPA
reproduces many unique features of cosmic gamma-ray bursts.
Hence the DRPA may shed important new light
on the unsolved
problem of gamma-ray bursts.
When a hot, strongly magnetized ($\beta$=thermal
pressure/plasma pressure $\le 1$) collisionless plasma
expands into a vacuum or low-density surrounding,
an electromagnetic pulse is formed at the
expansion surface, which traps and accelerates
a small fraction of surface particles via the
pondermotive force\cite{lian03,mill59},
and the magnetic energy is
efficiently converted into directed particle energy .

In our previous Letters\cite{lian03,nish03}, the DRPA
was studied by
performing
2-1/2D PIC simulations for
both 
electron-positron (mass ratio is $m_{i}/m_{e}=1$) and
electron-ion ($m_{i}/m_{e}=100$) plasmas with
a specific set of initial
parameters, and only carried out to 1000$\Omega_{e}^{-1}$
(where $\Omega_{e}=eB_{0}/m_{e}$ is the electron cyclotron
frequency defined by the initial magnetic field).
In this paper, we will study the parameter
dependence
and the long-term evolution of the DRPA 
of electron-positron plasmas by means of a
series of large-scale
PIC simulations.

The outline of this paper is as follows.
Section~\ref{sec:mode} describes the simulation model and
the simulation parameters.
The benchmark results of Liang {\it et al.}\cite{lian03}
are reviewed and summarized in
Section~\ref{subsec:diam} for comparison with other cases.
The long-term evolution of the DRPA
is described in Sec.~\ref{subsec:long}.
The initial parameter dependence of DRPA is
discussed in
Secs.~\ref{subsec:freq} and~\ref{subsec:plas}.
Section~\ref{sec:summ} gives a summary of our results.
%

\section{Simulation model}
\label{sec:mode}

In our particle simulation code\cite{nish02},
we use a 2-1/2D
explicit simulation scheme
based on the PIC method
for time advancing
of plasma particles.
In this method, spatial grids are
introduced to calculate the field quantities, and the
grid
separations are uniform, $\Delta x=\Delta z=\lambda
_{e}$,
where $\lambda _{e}$ is the electron Debye length defined
by $c/\omega_{pe}$
($c$ is the speed of light and the electron plasma
frequency
is $\omega_{pe}=\sqrt{e^{2}n_{0}/\epsilon _{0}m_{e}}$;
$e$ is elementary charge, $n_{0}$ is the initial electron
density, and $\epsilon _{0}$ is the dielectric constant
of
vacuum).
In our simulations, the thermal speed of electrons is
almost
equal to $c$ because the plasma temperature is
$k_{B}T_{e}=k_{B}T_{p}=5
MeV$, where the subscripts $e$ and $p$ refer to electrons
and positrons.
The simulation domain on the $x-z$ plane is
$-L_{x}/2\leq x\leq L_{x}/2$ and
$-L_{z}/2\leq z\leq L_{z}/2$.

We treat plasma particles
as superparticles, with 20 superparticles per cell
(if the superparticles are spread
uniformly on the entire
simulation domain)
for each component (electrons and positrons).
Most of the particle densities, however, concentrate on
the
expanding surface in our study and we are mainly
interested in particle behaviors near the
surface.
Thus, this number of superparticles is enough
to accurately resolve the physics of surface particle
acceleration.
To prevent large violations of Gauss's law due to
numerical noises,
the Marder's method for the electric field
correction\cite{mard87}
is adopted at every time step in the code.

We use a doubly periodic system in $x$ and $z$
directions,
and the system length is $L_{x}=3840\Delta x$ (Run A),
$3840\Delta x$ (Run B), $1920\Delta x$ (Run C), and
$7680\Delta x$ (run D)
and $L_{z}=
10\Delta z$ (for all four Runs).
Initially, the electron and
positron distributions are assumed
to be relativistic Maxwellian
with spatially uniform temperature,
$k_{B}T_{e}=k_{B}T_{p}=5MeV$.
The spatial distribution of initial plasmas is assumed to
be a uniform
slab with the lengths $12\Delta x\times 10\Delta
z$,(runs A,B,C) and $60\Delta x\times 10\Delta z$ (run D)
and the
plasmas are located in the center of the system. The
background magnetic field
$\mbox{\boldmath$B$}_{0}=(0,B_{0},
0)$ initially exists only inside the plasma.

We will first compare three simulation results
by changing the frequency ratio
$\omega_{pe}/\Omega_{e}$.
The values are
$\omega_{pe}/\Omega_{e}=0.104$ (Run A), 1.0 (Run B),
and 0.01 (Run C).
At the end we will also discuss the result of changing
the initial slab thickness (Run D).
%
\section{Results}
\label{sec:resu}

\subsection{Diamagnetic relativistic pulse accelerator}
\label{subsec:diam}
%
First, we illustrate the mechanism of the diamagnetic
relativistic
pulse accelerator (DRPA) by reviewing the
early ($t\Omega_{e}\le 10^{3}$) results of the benchmark
Run A (Liang {\it et al.}, 2003\cite{lian03}).
In Run A, the magnetic field energy is initially almost
equal to the thermal energy of plasma particles, i.e.,
electron and positron beta is $\beta_{e}=\beta_{p}=0.551$
($\beta=2\mu_{0}n_{0}k_{B}T/B_{0}^{2}$, where $\mu_{0}$
is
the magnetic permeability of vacuum).

\begin{figure}
\includegraphics[width=8cm]{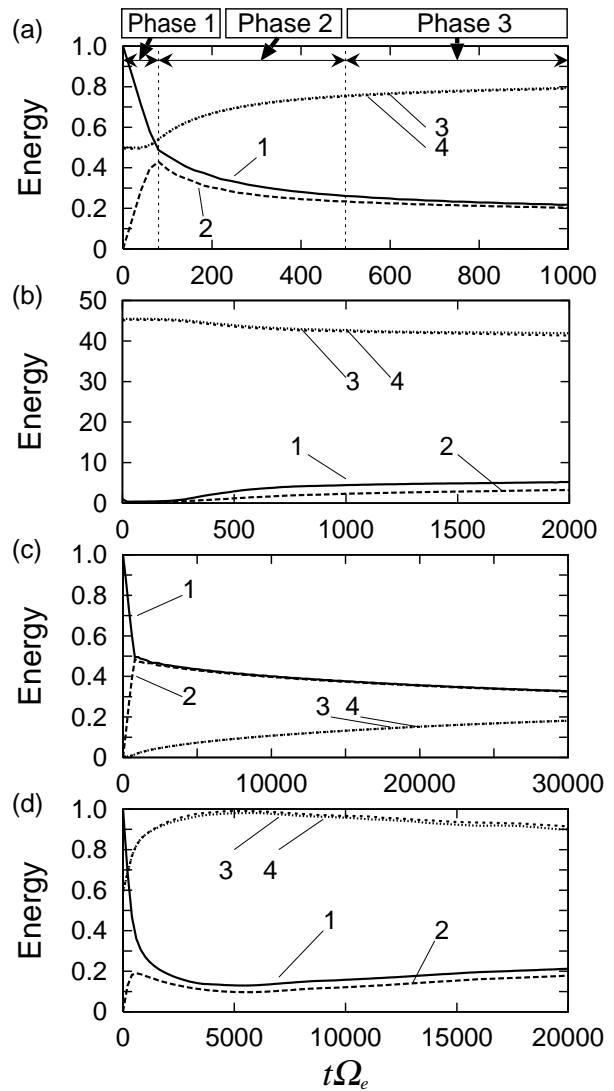}
\caption{\label{totenergy-30}
         System-integrated energy in the magnetic field (line 1),
         electric field (line 2), electrons (line 3), 
         and positrons (line 4) as
         functions
         of time for (a) Run A, (b) B, (c) C, and (d) D.
         Each energy component is normalized by the
         initial
         magnetic field energy in each figure.}
\end{figure}

In Figure~\ref{totenergy-30}(a),
the system-integrated energies of the magnetic field (line 1),
electric field (line 2), electrons (line 3), 
and positrons (line 4) are shown as
functions of time.
The initial static magnetic field
rapidly transforms into two oppositely propagating
electromagnetic pulses, which
propagate towards the vacuum region.
The plasma
simultaneously
begins to expand.
We see the rapid
decrease of the magnetic energy and the increase of
the electric field energy due to the generation of the
electromagnetic
(EM) pulse during the period shown as Phase 1
in
Fig.~\ref{totenergy-30}(a).
In Phase 2,
both the magnetic and electric field energies
begin to decrease
and they are converted into the electron and positron
kinetic
energies. Also,
the EM pulse with the amplitude $cB_{y}\sim
E_{z}$ is formed at the expansion surface during
this period.
In Phase 3, the energy conversion from the field energies
to the particle kinetic energies still continues although
the conversion rate becomes slower compared with that in
Phase 2.

\begin{figure}
\includegraphics[width=8cm]{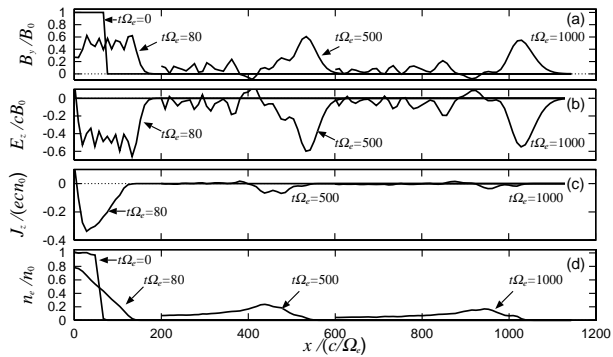}
\caption{\label{fields-x30}
         Spatial profiles of the (a) magnetic field
         $B_{y}$, (b) electric field $E_{z}$, (c) current
         density $J_{z}$, and (d) electron density
         $n_{e}$
         as functions of $x$ at $z=0$ for Run A.
         The quantities only in the $x\geq 0$ region are
         shown at $t\Omega_{e}=0$, 80, 500, and 1000
         (reproduced from Ref.[2]).}
\end{figure}
\begin{figure}
\includegraphics[width=8cm]{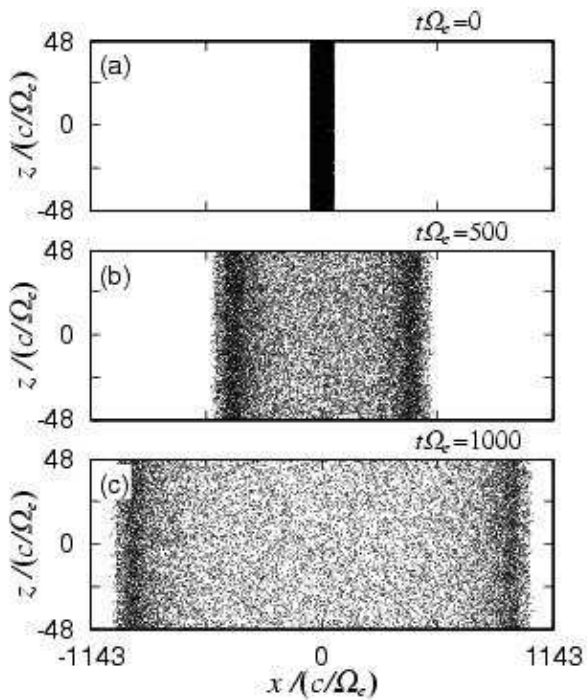}
\caption{\label{x-z30}
         2-D Electron distribution on the simulation
         plane
         for Run A at (a) $t\Omega_{e}=0$, (b) 500, and
         (c) 1000.}
\end{figure}
%
Figure~\ref{fields-x30} illustrates the spatial profiles
of the (a) magnetic field $B_{y}$, (b) electric field
$E_{z}$,
(c) current density $J_{z}$, and (d) electron density
$n_{e}$
at $t\Omega_{e}=0$, 80, 500, and 1000 for Run A.
The magnetic field and the electron density for
$x\leq 0$ are identical, on the other hand,
the electric field and the current density
become antisymmetric about $x=0$.
The values of $E_{z}$ and $J_{z}$ are zero
at $t\Omega_{e}=0$.
Only the profiles near the expansion surface are drawn
at $t\Omega_{e}=500$ and 1000 to avoid making
the figure cluttery.
At $t\Omega_{e}=80$,
the EM pulse
is formed
and begin to expand with plasmas into the vacuum.
The gradient drift in the z-direction induces a
current density.
At $t\Omega_{e}=500$, the EM pulse
has been formed at the expansion surface.
Most of the plasma particles follow the EM pulse
propagation and
the plasma density has a peak corresponding to particles
trapped by the pondermotive force of
the pulse.
The amplitude of the EM pulse is gradually damped
as EM energy is transformed into particle energy.

The $x-z$ configuration of electrons
is plotted at three different times in 
Figure~\ref{x-z30}.
The electron-positron slab with the internal magnetic
field expands rapidly into the vacuum region.
The behavior of positrons is identical with that of
electrons.
The expansion speed of the plasma surface in the $x$
direction
is almost
the speed of light, $V_{x}\simeq c$.
Several simulation studies of non-relativistic plasma expansion
across an initially uniform magnetic field have been 
carried out\cite{sydo83,wins88,sgro89,gisl89},
and the lower hybrid drift instability has been found
to deform the structure of the expanding plasma surface.
As shown in Fig.~\ref{x-z30}, there appears no
instability
at the plasma surface in our result.
There are two main differences between our simulation and
the previous studies\cite{sydo83,wins88,sgro89,gisl89}:
(a) in their studies, an initial magnetic field exists
uniformly all over the simulation plane and the plasma
expands (or diffuses) across the field,
(b) the electron-ion plasma was considered in their
studies. 
Thus, the charge separation due to the large
mass ratio causes the different drift speed between 
electrons and ions, which makes the lower hybrid drift
wave unstable.
These differences may explain the absence
of instability in our
simulation. In addition,
the effective Larmor radius of relativistic plasmas
given by $\gamma (c/\Omega_{e})$
is larger than our system length
in the $z$ direction at late times.
Thus, any effect of the Larmor radius play no
important role in our simulations, and we cannot see
any effect caused by the Larmor
motion of
relativistic plasmas.
%

\begin{figure}
\includegraphics[width=8cm]{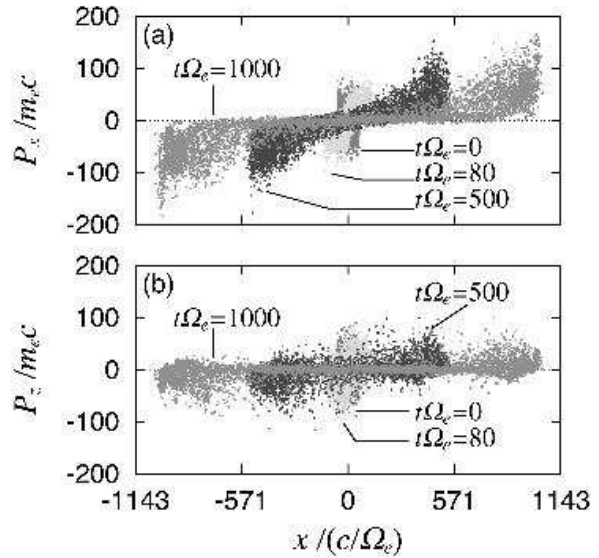}
\caption{\label{x-p30}
         Comparison of phase plots of electrons in the
         (a) $x-P_{x}$ and
         (b) $x-P_{z}$ spaces for Run A at
         $t\Omega_{e}=0$,
         80, 500, and 1000.}
\end{figure}
%
Figure~\ref{x-p30} shows phase plots of electrons at four
different times in the space (a) $x-P_{x}$ and (b)
$x-P_{z}$ for Run A.
It is found that most of the electron population moves to
the expansion surface and
the momentum $P_{x}$ of the surface
electrons increases as time elapses.
The averaged momentum reaches
$<P_{x}>\simeq 100m_{e}c$ and the maximum momentum
of energetic particles reaches $P_{x,max}\sim 200m_{e}c$
at $t\Omega_{e}=1000$.
On the other hand, the momentum $P_{z}$ indicates
no dramatic increase even at late times.
This means that the plasma particles are effectively
accelerated into the $x$ direction rather than the $z$
direction.

%
\subsection{Long-term evolution of the DRPA}
\label{subsec:long}

\begin{figure}
\includegraphics[width=9cm]{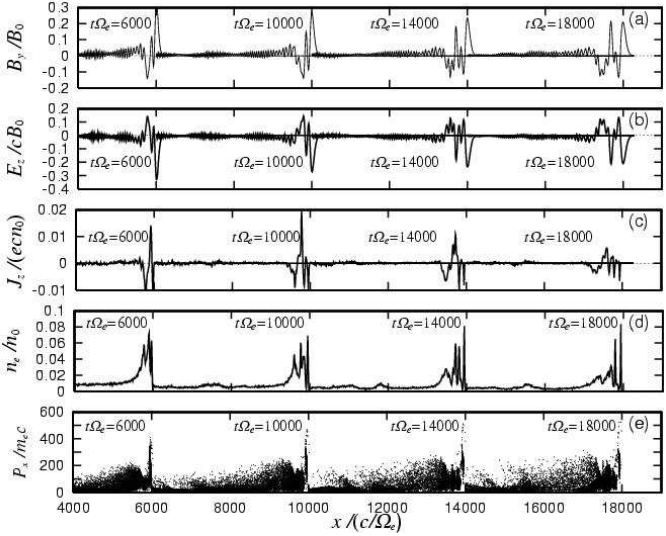}
\caption{\label{late30}
         Spatial profiles of the (a) magnetic field
         $B_{y}$, (b) electric field $E_{y}$, (c)
         current density $J_{z}$,
         and (d) electron density $n_{e}$ as functions of
         $x$ at $z=0$, and
         (e) phase plots of electrons in
         the $x-P_{x}$ space for Run A.
         The quantities only in the $x\geq 0$ region are
         shown at $t\Omega_{e}=6000$, 10000, 14000, and
         18000.}
\end{figure}

Let us focus on the long-term evolution of the DRPA in
Run A.
Figure~\ref{late30} shows the spatial profiles of the
(a) magnetic field $B_{y}$,
(b) electric field $E_{z}$,
(c) current density $J_{z}$,
(d) electron density
$n_{e}$,
and (e) the phase plots $x-P_{x}$ of electrons for Run A
at $t\Omega_{e}=6000$, 10000, 14000, and 18000.
Only the portion near the expansion front
at each time has been
displayed in each panel.
In Figs.~\ref{late30}(a) and ~\ref{late30}(b), 
the wave packet is dispersively
spreading and the EM
pulse amplitude gradually decreases as time elapses.
The reduction of the pulse amplitude corresponds to
the conversion of the EM energy into
the
kinetic energy of the expanding plasma.
In Fig.~\ref{late30}(e),
we recognize the electron momentum $P_{x}$
near the first pulse of the wave packet dramatically
increases as time goes on.
The maximum value of $P_{x}$ reaches more than
$500m_{e}c$
at $t\Omega_{e}=18000$.

Another significant feature of the late time behavior in
Run A
is the bifurcation of plasma density.
In the initial phase of the DRPA, the electron density
near
the wave packet has only one peak as shown
in Fig.~\ref{fields-x30}(d).
At late times in Fig.~\ref{late30}(d), the peak has been
split into several peaks. The bifurcation of
the plasma density profile
results from the development of
new transverse current and reverse currents behind the
front, which forms new traps for particles in the
pondermotive potential.

\begin{figure}
\includegraphics[width=8cm]{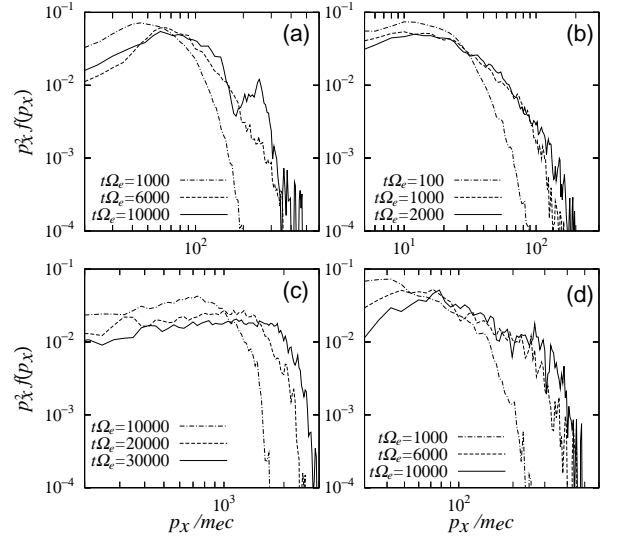}
\caption{\label{f(px)30}
         Electron momentum distributions for (a) Run A,
         (b) B, (c) C, and (d) D.
         Each distribution consists of the electrons near
         the
         electromagnetic pulse at $x\geq 0$.}
\end{figure}

\begin{figure}
\includegraphics[width=8cm]{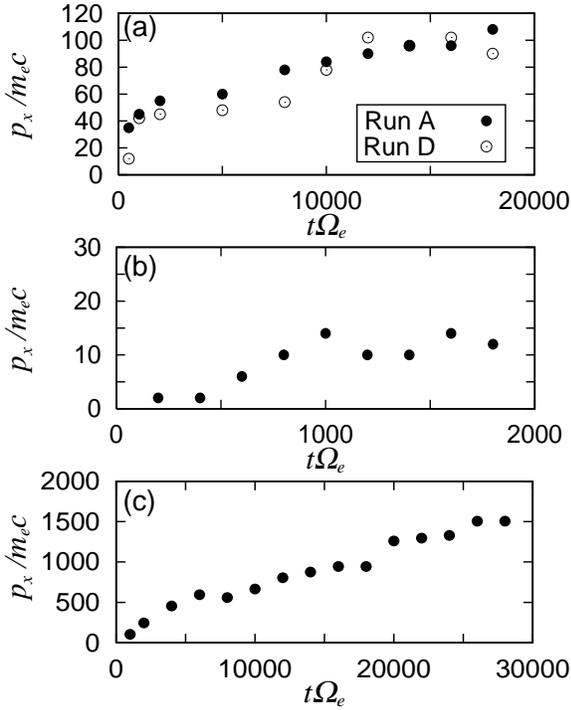}
\caption{\label{maxp-t}
         Temporal evolutions of the peak electron
         momentum at which
         the momentum distribution attains its maximum
         value for Runs
         (a) A and D, (b) B, and (c) C.}
\end{figure}

Figure~\ref{f(px)30}(a) shows the momentum distributions of
electrons which are located in the vicinity of the EM pulse
at $t\Omega_{e}=1000$, 6000, and 10000.
The momentum peak at which the distribution takes a maximum value
tends to increase as time elapses, i.e.,
$P_{x}\simeq 45m_{e}c$ at $t\Omega_{e}=1000$,
$P_{x}\simeq 60m_{e}c$ at $t\Omega_{e}=6000$, and
$P_{x}\simeq 80m_{e}c$ at $t\Omega_{e}=10000$
in Fig.~\ref{f(px)30}(a).
Also in Figure~\ref{maxp-t}(a), the temporal evolution of this
momentum is shown until $t\Omega_{e}=20000$ for Run A.
This temporal increment of the momentum verifies that the
average electon near the pulse is gradually accelerated.
In addition, we can see in Fig.~\ref{f(px)30}(a)
that the tail part of the distribution
develops into a quasi-power-law.
At $t\Omega_{e}=10000$,
a second peak of the distribution
appears at a high value $P_{x}$, which is
consisted of
non-thermal electrons.
In short, some of the surface electrons is more
accelerated than other
surface electrons.
This multi-peak structure of the momentum distribution
may be related to the bifurcation of the electron
density
shown in Fig.~\ref{late30}(d).

%
\begin{figure}
\includegraphics[width=9cm]{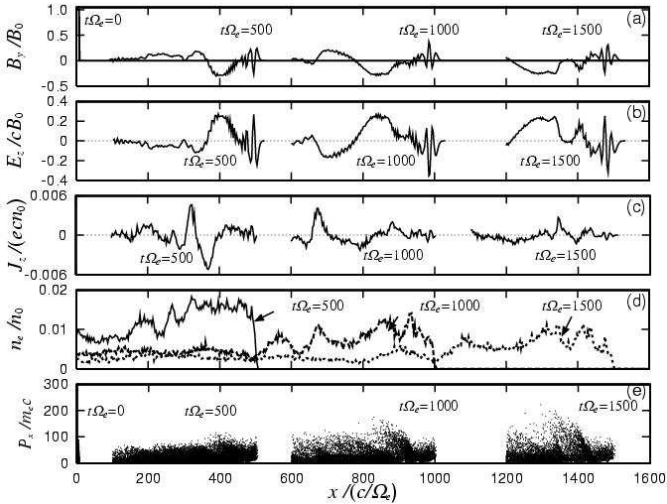}
\caption{\label{fld-x44}
         Spatial profiles of the (a) magnetic field
         $B_{y}$, (b) electric field $E_{y}$, (c)
         current density $J_{z}$,
         and (d) electron density $n_{e}$ as functions of
         $x$ at $z=0$, and
         (e) phase plots of electrons in
         the $x-P_{x}$ space for Run B.
         The quantities only in the $x\geq 0$ region are
         shown at $t\Omega_{e}=0$, 500, 1000, and
         1500.}
\end{figure}
%
\subsection{Frequency ratio dependence of the DRPA}
\label{subsec:freq}

In the previous sections, we summarize the results
of Run A (with the frequency ratio
$\omega_{pe}/\Omega_{e}=0.105$).
Next, we examine the
frequency ratio dependence of the DRPA.
The frequency ratios discussed here are
$\omega_{pe}/\Omega_{e}=1.0$ (Run B) and
$\omega_{pe}/\Omega_{e}=0.01$ (Run C).
Other fundamental simulation parameters are exactly same
as those of Run A.

Figure~\ref{fld-x44} displays the spatial profiles
of the same quantities shown in Fig.~\ref{fields-x30},
in the case of Run B, at different times.
Figs.~\ref{fld-x44}(a), (b), (c), and (e)
mainly show the region near the EM pulse at each
time, and the values of $E_{z}$ and $J_{z}$ are zero at
$t\Omega_{e}=0$.
The electron density $n_{e}$ at $t\Omega_{e}=0$ is not
shown in Fig.~\ref{fld-x44}(d) since
the initial density $n_{e}/n_{0}=1$ is much larger
than the values at the other three times.
Compared with the results of Run A, no apparent EM pulse
with large amplitudes can be formed at the expansion
surface in Figs.~\ref{fld-x44}(a) and ~\ref{fld-x44}(b).
In Fig.~\ref{fld-x44}(d), we find that the electrons are
distributed in a broad region, rather than confined in
the EM pulse.
We see no sign of bifurcation of the density pulse
like in Run A.
In Run B, initially the thermal energy of plasma
particles
is much larger than the magnetic field energy
because of
$\beta_{e}=\beta_{p}=50$.
Temporal evolution of each energy component is shown 
in Fig.~\ref{totenergy-30}(b).
In Run A, the field energy is converted into the kinetic
energy of surface particles by the DRPA,
thus the total field energy decreases as time elapses.
In Run B, however, the field energy continues to increase
as shown in Fig.~\ref{totenergy-30}(b).
This increase of the EM field energy is mainly
caused by the field fluctuation in the regions away
from the expansion surface.

\begin{figure}
\includegraphics[width=8cm]{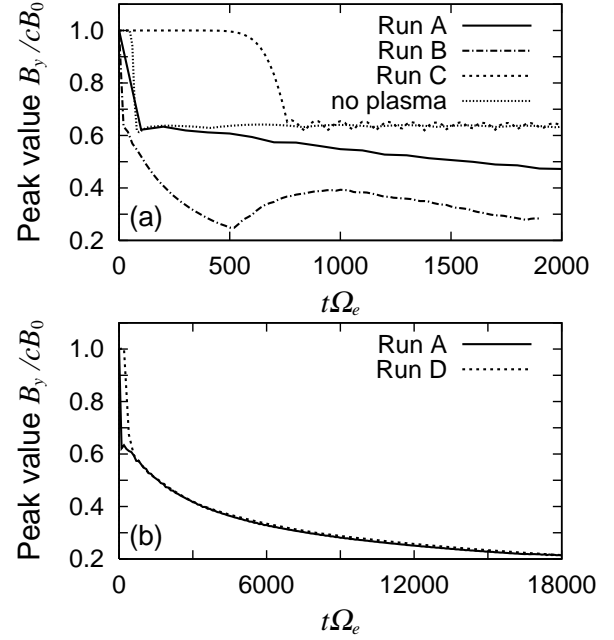}
\caption{\label{peakB-t}
         Temporal evolutions of the peak value of the
         magnetic
         field of the pulse for (a) Runs A, B, C, and
         the case with
         no plasma (pure vacuum EM pulse), and (b)
         comparison between Runs A and D.}
\end{figure}
%
We focus on the amplitude of the propagating magnetic field 
formed at the expansion surface.
Temporal evolution of the peak value of the magnetic
field
pulse $B_{y}$ is shown in Figure~\ref{peakB-t}(a).
As for Run A,
the peak value dramatically decreases during
$0<t\Omega_{e}<80$
(this period corresponds to Phase 1 in
Fig.~\ref{totenergy-30})
and then continues to decrease at a constant damping rate
(this period corresponds to Phases 2 and 3
in Fig.~\ref{totenergy-30}).
We can see the similar tendency in the temporal evolution
of
the peak value for Run B
during $0<t\Omega_{e}<500$, 
but the damping rate of the
peak value
is much larger than that of Run A.
At $t\Omega_{e}\simeq 500$, the peak value becomes
$B_{y}<0.3B_{0}$.
This means that the magnetic field energy in Run B
is dissipated
much faster than that in Run A.
Although the magnetic field energy is rapidly converted
into particle energy, the total kinetic energy of Run B
hardly changes since the initial plasma beta is very
high.
Therefore, the evolution can be
described as the free-expansion of the
plasma rather than
the expansion accompanied by the
acceleration of the EM pulse.
The electron momentum peak of Run B, at which the momentum
distribution
takes a maximum, is shown as a function of time in
Fig.~\ref{maxp-t}(b). The momentum remains around
$P_{x}\simeq 15m_{e}c$ even when the amplitude of the
magnetic
field pulse becomes $B_{y}\simeq 0.3B_{0}$ at
$t\Omega_{e}=500$
because the converted magnetic field energy is too small to
accelerate all the plasma particles
via the DRPA. Thus, the
electrons 
at late times exhibit almost uniform distribution 
by the free-expansion and the
density pulse is not formed in Run B, as shown
in Fig.~\ref{fld-x44}(d).
We see the evolution of the electron momentum distribution
for Run B in Fig.~\ref{f(px)30}(b).
The distribution keeps a Maxwellian form even at late times.
This means that the plasma is consisted almost only of
the thermal component.
%

\begin{figure}
\includegraphics[width=9cm]{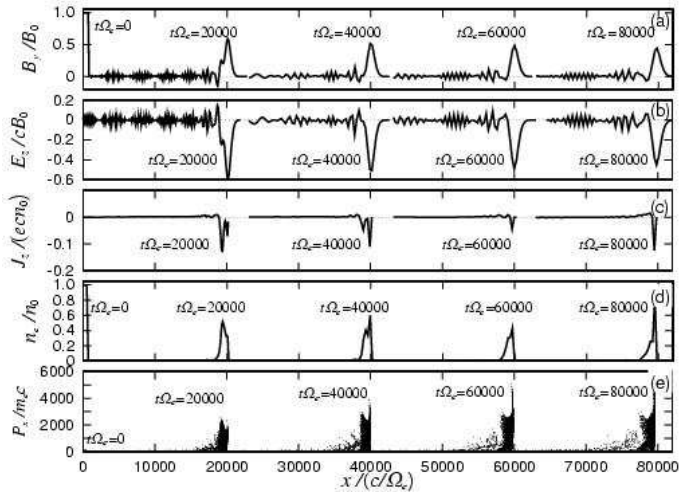}
\caption{\label{fld-x34}
         Spatial profiles of the (a) magnetic field
         $B_{y}$, (b) electric field $E_{y}$, (c)
         current density $J_{z}$,
         and (d) electron density $n_{e}$ as functions of
         $x$ at $z=0$, and
         (e) phase plots of electrons in
         the $x-P_{x}$ space for Run C.
         The quantities only in the $x\geq 0$ region are
         shown at $t\Omega_{e}=0$, 20000, 40000, 60000, and
         80000.}
\end{figure}
%
Figure~\ref{fld-x34} shows the results of Run C
($\omega_{pe}/\Omega_{e}=0.01$ and
$\beta_{e}=\beta_{p}=0.005$).
In this run the initial
magnetic field energy is much greater than the thermal
energy
of plasma particles.
The converted energy of the EM pulse can be effectively
used for the particle acceleration
[see Fig.~\ref{totenergy-30}(c)].
In Fig.~\ref{fld-x34}(d), most of the electrons has been
accelerated owing to the EM pulse and the electron
density is much more concentrated at the expanding
surface.
Temporal evolution of the maximum amplitude of the pulse
magnetic field in Run C is also plotted in
Fig.~\ref{peakB-t}(a).
The period during $0<t\Omega_{e}<750$ represents an
initial relaxation phase in Run C,
which corresponds to Phase 1 in Fig.~\ref{totenergy-30}(a)
of
Run A.
After that, the peak value stays almost constant,
$B_{y}\simeq 0.62B_{0}$.
For comparison, the result of a vacuum EM pulse
without plasma is shown in Fig.~\ref{peakB-t}(a).
In this run without plasma, we set up the initial
magnetic
field $B_{y}$ to be $B_{y}(x)=B_{0}$ for $|x|<3\Delta x$ and
$B_{y}(x)=0$ for $|x|\geq 3\Delta x$, and the electric field
is initially zero everywhere.
The EM pulse generated by the spatial gradient of the
initial
magnetic field propagates according to the wave equation
in a vacuum.
It is found that the amplitude of the EM wave is almost
unchanged after the initial relaxation phase
($0<t\Omega_{e}<80$) in Fig.~\ref{peakB-t}(a).
The EM pulse in Run C is very similar to the
pulse without plasma since the initial plasma beta is
small, $\beta=0.005$, and only a small portion of the
total magnetic field energy is converted into the
plasma kinetic energy.

The evolution of the electron momentum distribution
for Run C is shown in Fig.~\ref{f(px)30}(c).
Since most of electrons are accelerated by the DRPA,
the tail part of the distribution exhibits a power-law
with sharp slope.
In Fig.~\ref{maxp-t}(c), the peak momentum
of the surface electrons
increases rapidly as time elapses, compared to Runs A and
B
with higher frequency ratios.
The average momentum of electrons reaches
$<P_{x}>\simeq 1500m_{e}c$ at $t\Omega_{e}=30000$.

%
\begin{figure}
\includegraphics[width=9cm]{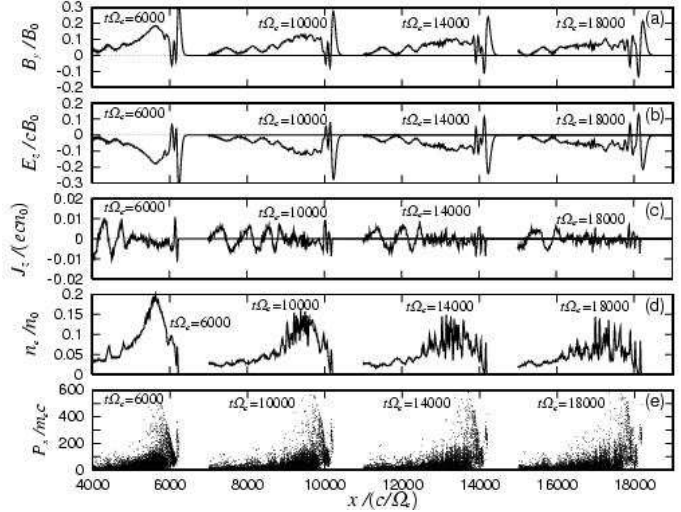}
\caption{\label{fld-x30e}
         Spatial profiles of the (a) magnetic field
         $B_{y}$, (b) electric field $E_{y}$, (c)
         current density $J_{z}$,
         and (d) electron density $n_{e}$ as functions of
         $x$ at $z=0$, and
         (e) phase plots of electrons in
         the $x-P_{x}$ space for Run D.
         The quantities only in the $x\geq 0$ region are
         shown at $t\Omega_{e}=0$, 6000, 10000, 14000, and
         18000.}
\end{figure}
\subsection{Dependence on the initial plasma width}
\label{subsec:plas}
%
In this section,
we examine how the width of the initial plasma
can affect the DRPA.
The widths of the density pulse of expanding plasmas tend
to scale with the width of
the initial plasma, as we expect from causality.
The propagating EM pulses have the same tendency since
the
initial magnetic field has the same width as the plasma
density.
We have carried out Run D, in which the frequency ratio
is
$\omega_{pe}/\Omega_{e}=0.105$ and the initial width of
plasma is five times that of Run A.

The temporal evolution of each energy is shown in
Fig.~\ref{totenergy-30}(d).
The apparent energy conversion from field energies into
particle energies due to the DRPA can be seen until
$t\Omega_{e}\simeq 5000$, which is similar to the
evolution in Fig.~\ref{totenergy-30}(a).
After that time, the total EM energy slightly increases
and the total particle energies tends to decrease.
These slight energy changes against the DRPA is
caused by the local energy changes in the regions
away from the expansion surface.
In those regions, the particle acceleration due to
the DRPA hardly works efficiently.
Even though the total particle energies slightly decreases,
we will show that the plasmas near the propagating
EM pulse continue to be energized after 
$t\Omega_{e}\simeq 5000$.

The electron momentum distributions for Run D
are displayed in Fig.~\ref{f(px)30}(d) at three
different times.
As time elapses, the average momentum of surface
electrons increases, and we recognize
a lot of peaks in the high momentum part at
$t\Omega_{e}=10000$.
As we will show later in Figure~\ref{fld-x30e},
these peaks correspond to a repeated bifurcation
of electron density at the expansion surface.
In Fig.~\ref{maxp-t}(a), the results of Run D are also
plotted.
There is no big difference between the peak electron
momenta of Runs A and D, except in D the acceleration
seems more delayed due to the prolific early
bifurcations.
However the maximum energy achieved by the most energetic
particles is higher in D than in A.  This is likely
caused by the larger coherence length for acceleration in
D due to the wider EM pulse.

Fig.~\ref{fld-x30e} shows the spatial profiles of the
(a) magnetic field $B_{y}$ and (b) electron density
$n_{e}$,
and (c) the phase plots $x-P_{x}$ of electrons for Run D
at
$t\Omega_{e}=6000$, 10000, 14000, and 18000.
Only the portion near the expansion surface at each time
has been displayed in each panel.
Compared with the electron density of Run A in
Fig.~\ref{late30}(b), the width of the density of the
expanding
electrons $w$ becomes larger, e.g., $w\simeq
2000c/\Omega_{e}$
at $t\Omega_{e}=18000$ in Fig.~\ref{fld-x30e}(b).
We also recognize many more bifurcations of the
electron density
at late times.

%
\section{Summary}
\label{sec:summ}
%
We have used two-and-a-half-dimensional large-scale 
particle-in-cell
plasma simulations to investigate the particle
energization
of expanding relativistic electron-positron plasmas.
When magnetized plasmas with high temperature expand into
a vacuum, the particles at the expansion surface are
accelerated in the expansion direction.
This energization mechanism is called
DRPA (diamagnetic relativistic pulse
accelerator)\cite{lian03}.

In this paper, we mainly discuss the long-term
evolution and the initial
parameter dependences of DRPA.
In the case of the electron-positron plasma with the
frequency
ratio $\omega_{pe}/\Omega_{e}=0.105$ and the temperature
$k_{B}T_{e}=k_{B}T_{p}=5MeV$, an electromagnetic (EM)
pulse
with large amplitude is generated and propagates
into a
vacuum. The plasma particles are
trapped and accelerated by the
pondermotive force. As time elapses, since the magnetic
field energy is converted into the kinetic energy of the
particles, the amplitude of the EM pulse is decreased and
the
particles near the expansion surface are energized.
The expanding plasma density is concentrated on the
vicinity
of the expansion surface at the early phase of the
expansion.
And at the late phase, we observe the
bifurcation of plasma
density due to the formation of new transverse currents.

The properties of DRPA strongly depend on the
frequency ratio
$\omega_{pe}/\Omega_{e}$.
In the case of a high frequency ratio
($\omega_{pe}/\Omega_{e}=1.0$), the EM pulse is rapidly
damped and the plasma is
expanding freely with little acceleration since
the energy of the plasma is much larger than the
magnetic energy.
On the other hand, in the case of a low frequency ratio
($\omega_{pe}/\Omega_{e}=0.01$),
the magnetic
energy is
much higher than the plasma thermal energy.
Most of the expanding plasma is trapped by the
EM pulse and the plasma is efficiently
energized at late times, achieving higher Lorentz factors
than in the other cases.

In addition to the frequency ratio, we have studied
the effect of varying the
initial width of the plasma slab.
As expected, the width of the asymptotic EM pulse tends
to be proportional to
the initial width of the plasma slab.
Both the EM pulse and the
plasma density
are broadened, and the density pulse develops
many more peaks
at late times due to prolific bifurcations.
We also find that the maximum
energy of the accelerated particles is higher 
in this case due to the
longer pulse length.

Finally, we note the system length in the $z$ direction
in our two-dimensional simulation domain.
As we mentioned in Sec.~\ref{subsec:diam},
the effective Larmor radius of relativistic electron-positron
plasmas $\gamma (c/\Omega_{e})$ becomes comparable or 
larger than
$L_{z}$ at late times.
Thus, the present system length $L_{z}$ will be
insufficient to fully cover the physical phenomena scaled
by the Larmor radius if there exist some instabilities
which can develop along $L_{z}$ on that scale.
However, to our knowledge, previously observed
instabilities in non-relativistic 
simulations\cite{sydo83,wins88,sgro89}, 
such as the lower hybrid drift instability,
do not grow in our electron-positron relativistic
plasmas.
Future advanced simulations 
will be able to give
the clear answer to this problem through 
further simulation studies with larger system lengths.

%
\acknowledgments

The work of K.N. was performed under the auspices of the
U. S. DOE
and was supported by the DOE Office of
Basic
Energy Sciences, Division of Engineering and Geosciences,
the
LDRD Program at
Los Alamos,
and the Sun-Earth Connections Theory Program of NASA.
E.L. was partially supported by NASA grant NAG5-7980 and
LLNL contract B510243.



%
\end{document}